\documentclass[10pt]{IEEEtran}

\usepackage{float}
\usepackage{cite}
\usepackage{amsbsy}
\usepackage{amssymb}
\usepackage{amscd}
\usepackage{amsmath}

\newcommand\pp[1]{{{\left(#1\right)}}}

\usepackage{dot2texi,tikz,pgfplots}
\usetikzlibrary{shapes,arrows}
\usetikzlibrary{positioning}
\usetikzlibrary{math}
\tikzset{main node/.style={circle,draw},
            }

\newcommand{\singleantenna}[2]{
\draw[very thick,-] (#1,#2-0.1) -- (#1,#2+.05);
\draw[very thick,-] (#1,#2+.05) -- (#1-.135,#2+.175);
\draw[very thick,-] (#1,#2+.05) -- (#1+.135,#2+.175);
\draw[very thick,-] (#1,#2-0.1) -- (#1-0.3,#2-0.1);
}

\usepackage[color=black,opacity=1]{background}

\SetBgContents{ \begin{minipage}{35cm}{\copyright 2018 IEEE. Personal use
      of this material is permitted. Permission from IEEE must be obtained
       for all other uses, in any current or future media, including
        reprinting/republishing this material for advertising or promotional purposes,
         creating new collective works, for resale or redistribution to servers or lists, 
         or reuse of any copyrighted component of this work in other works. This paper will appear in the IEEE  Wireless Communications Letters, 2018, DOI: 10.1109/LWC.2018.2802519.}\end{minipage}}
\SetBgScale{.5}
\SetBgAngle{0}
\SetBgPosition{current page.south west}
\SetBgHshift{20cm}
\SetBgVshift{2cm}

\title{Out-of-Band Radiation from Antenna Arrays Clarified}

\author{Erik G. Larsson and Liesbet Van der Perre\thanks{E. G. Larsson
    is with the Dept.\ of Electrical Engineering (ISY), Link\"oping
    University, 581 83 Link\"oping, Sweden. Email:
    \texttt{erik.g.larsson@liu.se}. Phone:
    +46-13-281312.}\thanks{L. V. d. Perre is with the Dept. of Electrical
    Engineering, KU Leuven, 3001 Leuven, Belgium.}\thanks{This work was
    supported in part by the Swedish Research Council (VR) and
    ELLIIT.}}

\begin{document}
\maketitle
\begin{abstract}
  Non-linearities in radio-frequency (RF) transceiver hardware,
  particularly in power amplifiers, cause distortion in-band and
  out-of-band.  Contrary to claims made in recent literature, in a
  multiple-antenna system this distortion is correlated across the
  antennas in the array. A significant implication of this fact is
  that out-of-band emissions caused by non-linearities are beamformed,
  in some cases into the same direction as the useful signal.
\end{abstract}

\section{Introduction}

Non-linearities in RF amplifiers and other electronics in wireless
transmitters give rise to unwanted emissions. A sinusoidal input with
(angular) frequency $\omega$ that passes through a non-linearity will
contain harmonic components with frequencies
$\omega,2\omega,3\omega,...$; see, for example, \cite{horlin2008}.
Two-tone waveforms consisting of two superimposed sinusoids with
frequencies $\omega$ and $\omega'$ will have spectral peaks at
$\omega$, $\omega'$, $\omega+\omega'$,
$\omega-\omega'$, $\omega'-\omega$,
$2\omega-\omega'$, $2\omega'-\omega$, $2\omega+\omega'$,
$2\omega'+\omega$, among others.  Signals with a general, but
band-limited, spectrum will see a widening of their bandwidth --
resulting in in-band distortion inside of their nominally occupied
frequency interval, as well as out-of-band distortion in adjacent
frequency bands.

In-band distortion is typically quantified in terms of an error vector
magnitude (EVM). This distortion translates directly into a quality
loss in the demodulated signal, and this loss can be substantial.
Out-of-band distortion represents, in many cases, an even more
significant concern. A typical constraint is that the relative power
radiated into the adjacent band shall be below a given threshold,
known as the adjacent channel leakage ratio (ACLR). For example, for
the LTE standard the ACLR must be below $-45$ dB. For waveforms with
high peak-to-average power ratios (notably, OFDM and certain multiuser-MIMO
precoded signals), large amplifier backoffs are required to meet
specified EVM and ACLR requirements.

Recently several papers have attempted to assess the impact of
hardware imperfections in transmitters equipped with multiple-antenna
arrays \cite{bjornson2014massive,zhang2015impact}, especially in the
context of Massive MIMO -- the leading 5G physical layer technology
that exploits large antenna arrays at base stations
\cite{MLYN2016book}.  One question of concern has been whether the
distortion created by hardware imperfections averages out or not, when
adding more antennas -- in a similar way as thermal noise averages
out.  Another, specific question of interest has been whether or not
the out-of-band emissions resulting from non-linearities have spatial
directivity
\cite{mollen2016out,zou2015impact,blandino2017analysis,gustavsson2014impact}.
As we show in this letter, there are serious issues both with the
models used, and the subsequent conclusions drawn, in several of these
papers (including some by the authors).

\textbf{Contribution:} This letter has two specific objectives.
First, we explain, from rigorous first principles, why and how the
distortion resulting from non-linearities is beamformed. Second, we
rectify various inaccurate statements made in recent literature.

While a mathematical analysis of non-linearities in general tends to
become rather intricate, our derivations rely only on elementary
manipulations of trigonometric identities. We hope that this will
render the exposition accessible to a wider readership.  We also
discuss the implications in single-user and multiple-user
communication scenarios exploiting antenna arrays.
  
\section{Two-Tone Waveform Through a Third-Order Non-Linearity}

Consider an array of antennas (indexed by $m$) used for beamformed
transmission.  The signals fed to the antennas have undergone
amplification in a device with non-linear behavior generating harmonics.
The question is whether the distortion caused by these non-linearities
adds up constructively or not, and if so, in what spatial directions.
 We assume that amplification devices with identical
characteristics are used on all antennas.\footnote{Measurements have validated that
 small differences among the amplifiers cannot be counted on to de-correlate the distortion.} 

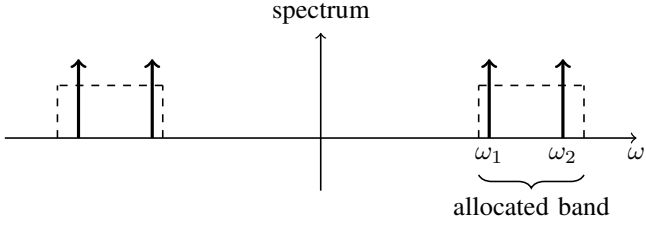
\begin{figure}[t!]
  \begin{center}
  \begin{tikzpicture}[scale=0.7]
  \node at (3.2,0) [anchor=north] {$\omega_1$};
  \node at (4.6,0) [anchor=north] {$\omega_2$};
  \draw[->,semithick] (-6,0) -- (6,0);
  \draw[->,semithick] (0,-1) -- (0,2);
  \node at (6,0) [anchor=north] {$\omega$};
  \node at (0,2) [anchor=south] {spectrum};
  \draw[-,semithick,dashed] (3,0) -- (3,1);
  \draw[-,semithick,dashed] (5,0) -- (5,1);
  \draw[-,semithick,dashed] (3,1) -- (5,1);
  \draw[-,semithick,dashed] (-3,0) -- (-3,1);
  \draw[-,semithick,dashed] (-5,0) -- (-5,1);
  \draw[-,semithick,dashed] (-3,1) -- (-5,1);
\draw[->,very thick] (3.2,0) -- (3.2,1.5);
\draw[->,very thick] (4.6,0) -- (4.6,1.5);
\draw[->,very thick] (-3.2,0) -- (-3.2,1.5);
\draw[->,very thick] (-4.6,0) -- (-4.6,1.5);

\draw[semithick,decoration={brace,amplitude=5pt,raise=0pt,mirror},decorate] (3,-.7) -- node[pos=0.5, inner sep=8pt, anchor=north] {allocated band} (5,-.7);   
  \end{tikzpicture}
  \end{center}
\vspace*{-.5cm}  \caption{\label{fig:band}Two-tone model, as representation of a signal in a frequency interval
allocated to a particular transmission.}
\end{figure}

To illustrate the principal phenomena, 
we assume that the $m$th amplifier is fed with a two-tone signal of the form,
\begin{align}
x_m(t)=  \cos(\omega_1 t + \phi_1^m)   + \cos(\omega_2 t + \phi_2^m).
\end{align}
This is a rather simple model, but it is sufficient to prove our
points (and commonly used in the analysis of non-linearities).  The
analysis is agnostic of $\omega_1$ and $\omega_2$, but in a practical
application both would belong to a frequency interval allocated to the
transmission of a useful signal (Figure~\ref{fig:band}).  The signal
$x_m(t)$ passes through a third-order memoryless
non-linearity,\footnote{This  third-order model is relevant for many
hardware components in practice. Note that second-order non-linearities are of no
  interest. With a second order non-linearity, $f(x)=x+\alpha x^2$, it
  can be verified that the output, after removal of frequency
  components far away from $\omega_1$ and $\omega_2$, and terms that
  constitute a scaled version of the input, is zero.}
\begin{align}\label{eq:thirdorder}
  f(x)=x+\alpha x^3.
\end{align}
The output signal is
\begin{align}\label{eq:1}
y_m(t)  & = \cos (\omega_1 t+\phi_1^m) 
  +  \cos (\omega_2 t+\phi_2^m)  \nonumber \\
  &\quad +  \alpha \pp{\cos (\omega_1 t+\phi_1^m) + \cos (\omega_2 t+\phi_2^m) }^3.
  \end{align}
Through the use of standard trigonometric identities, (\ref{eq:1}) can be
equivalently  written as
\begin{align} 
y_m(t) & = \pp{1+\frac{9\alpha}{4}}  \cos (\omega_1 t+\phi_1^m)  \nonumber \\
&\quad +  \pp{1+\frac{9\alpha}{4}}  \cos (\omega_2 t+\phi_2^m)  \nonumber \\
&\quad +\frac{3\alpha}{4}  \cos (2 \omega_2  t+\omega_1 t+2 \phi_2^m+\phi_1^m)  \nonumber \\
&\quad +\frac{3\alpha}{4} \cos (2 \omega_2 t-\omega_1 t+2  \phi_2^m-\phi_1^m)  \nonumber \\
&\quad +\frac{3\alpha}{4} \cos (\omega_2 t+2 \omega_1 t+\phi_2^m+2 \phi_1^m)  \nonumber \\
&\quad +\frac{3\alpha}{4} \cos (\omega_2 t-2 \omega_1 t+\phi_2^m-2 \phi_1^m)  \nonumber \\
&\quad +\frac{\alpha}{4} \cos (3 \omega_1 t+3 \phi_1^m )  \nonumber \\
&\quad +\frac{\alpha}{4} \cos (3 \omega_2 t+3 \phi_2^m).\label{eq:exp} 
\end{align}
 The first two
terms represent the desired signal, scaled by a constant. The remaining
 terms contain distortion created by the non-linearity.  The two
relevant distortion terms are those with frequencies near $\omega_1$
and $\omega_2$, that is,
\begin{align}
 &  \frac{3\alpha}{4}\cos (2 \omega_2 t-\omega_1 t+2  \phi_2^m-\phi_1^m) \label{eq:d1} \\
\mbox{and}\qquad &\frac{3  \alpha}{4} \cos (\omega_2 t-2 \omega_1 t+\phi_2^m-2 \phi_1^m). \label{eq:d2}
\end{align}  
All other terms are irrelevant, as they have frequencies far away from
$\omega_1$ and $\omega_2$ and in practice they would be suppressed by
the transmit antenna frequency response and therefore effectively not
transmitted.

Hence the focus now is on the terms (\ref{eq:d1}) and (\ref{eq:d2}),
and on the question whether they add up constructively or not in
particular spatial directions. To proceed with the analysis, we
consider a setup with two antennas and a receiver located in
the far field of the array. Then $m=1,2$.

\begin{figure}[t!]
  \begin{center}
  \begin{tikzpicture}[scale=0.8]
  \singleantenna{-0.3}{-1}
  \singleantenna{-0.3}{1}
  \node at (-0.7,-1) [anchor=east] {$\cos(\omega t + \phi^1)$};
  \node at (-0.7,1) [anchor=east]  {$\cos(\omega t + \phi^2)$};
  \draw[->,very thick] (0,1) -- (1,2);
  \draw[->,very thick] (0,-1) -- (2,1);
  \draw[dotted,very thick] (0,1) -- (1,0);
\draw[semithick,decoration={brace,amplitude=3pt,raise=5pt,mirror},decorate] (0,-1) -- node[pos=0.5, inner sep=5pt, anchor=north west] {propagation delay $\tau=\frac{\phi^2-\phi^1}{\omega}$ } (1,0);   
  \end{tikzpicture}
  \end{center}
  \caption{\label{fig:1}Two antennas that radiate a sinusoidal signal
    with (angular) frequency $\omega$.  If the phase difference of the
    sinusoids is $\phi^2-\phi^1$, then they combine constructively in
    the directions associated with a propagation delay of
    $(\phi^2-\phi^1 + n 2\pi)/\omega$ seconds, where $n$ is an
    integer.  If the antenna spacing is no more than half a
    wavelength, or equivalently the propagation delay corresponding to
    the antenna spacing is less than or equal to $\pi/\omega$, then
    there are no grating lobes and $\tau$ uniquely defines a single
    spatial direction.}
\end{figure}
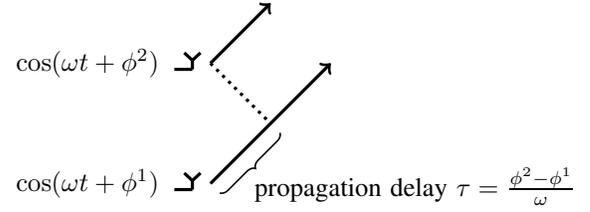

\section{Single-user beamforming case}\label{sec:SU}

We consider first the case of single-user beamforming.  Then the two
terms in (\ref{eq:exp}) that constitute the desired signal are
destined into the same spatial direction.  Let this spatial direction
be defined in terms of the difference in propagation delay, say
$\tau$, between the path from the receiver to antenna 2, respectively
from the receiver to antenna 1. Figure~\ref{fig:1} explains the
definition of $\tau$.

The two terms in (\ref{eq:exp}) that represent the desired signal add
up constructively in that very direction exactly if antenna 2 has a
phase lead over antenna 1 of $\omega_1\tau$ for the first sinusoid and
a phase lead of $\omega_2\tau$ for the second sinusoid, that
is,\footnote{All equations involving phase angles are modulo $2\pi$.}
\begin{align}
  \phi_1^2-\phi_1^1 & = \omega_1\tau, \label{eq:s1} \\
  \phi_2^2-\phi_2^1 & = \omega_2\tau. \label{eq:s2}
\end{align}
Suppose that the relative phases $\phi_1^2-\phi_1^1$ and
$\phi_2^2-\phi_2^1$ are chosen such that (\ref{eq:s1})--(\ref{eq:s2})
are satisfied. (The absolute phases $\phi_1^1$ and $\phi_2^1$ are
unimportant.)  Then, a direct calculation from
(\ref{eq:s1})--(\ref{eq:s2}) shows that,
\begin{align}
 (2\phi_2^2-\phi_1^2) - (2\phi_2^1-\phi_1^1) =
  (2\omega_2-\omega_1)\tau.
\end{align}
Consider now the first distortion term, (\ref{eq:d1}), which is a
sinusoid with frequency $ (2\omega_2-\omega_1)$. For this term,
antenna 2 will have a phase lead of $ (2\omega_2-\omega_1)\tau$ over
antenna 1. This means that the distortion term (\ref{eq:d1}) adds up
constructively in the direction defined by $\tau$ -- that is, the same
direction as into which the useful signal is beamformed.

A similar calculation yields the same conclusion for the second
distortion term, (\ref{eq:d2}). Note that with our model, if the
frequencies $2\omega_2-\omega_1$ and $\omega_2-2\omega_1$ lie inside
the band, the distortion terms impacts the EVM, otherwise they impact
the ACLR.

Generally, beamforming with a linear uniform antenna array in a
specific spatial direction may be achieved using either a conventional
analog phased array, or a digital implementation of such an array. 
The spatial directivity of the out-of-band distortion\footnote{Except
  for a small extra attenuation in the antenna chain of out-of-band
  signals close to the in-band.} will in the far-field have the same
radiation pattern, regardless of the actual implementation of the
beamforming. (See, for example, \cite{Balanis:2005}, for fundamental
theory of radiation patterns of antenna arrays.)  Considering typical
radiation patterns of arrays, the presence of out-of-band distortion is
not limited to one specific direction. This distortion will be present
in the entire beam within which the gain is significant, as well as in
the associated sidelobes. Pattern characteristics of harmonic and
intermodulation products in broad-band active phased arrays 
have been investigated and reported on in 
the past \cite{Sandrin1973, Hemmi2002}.
 
Many millimeter wave systems foreseen for 5G will use antenna arrays
to improve the link budget, and the above analysis applies to these
systems.  It is likely that the power amplifiers in such systems will
be operated in the non-linear region \cite{zou2016impact}.  Hence, the
presence of out-of-band radiation cannot be neglected.

In sub-6 GHz massive MIMO systems a similar effect will be experienced
when serving one single user.  Due to the randomness of path losses and
shadow fading, the effective channel gain typically differs
substantially between different terminals.  When power control is
applied to compensate for these differences and equalize the
quality-of-service to the terminals, most of the radiated power is
typically spent on only one of the terminals
\cite[p.~105]{MLYN2016book}. Hence, the effect will be seen with high
probability also in a multi-user scenario.
However, in general, for systems operating at lower frequencies in
multi-path conditions, the channel-matched beamforming will randomize
the non-linear distortion terms.  This fact has been established
through simulations in previously published papers; see, e.g.,
\cite{mollen2016out}.

The effect on the EVM and ACLR is as follows (for the single-user case):
\begin{itemize}
\item The EVM respectively ACLR values measured \emph{at the
  transmitter}, at any of the antenna ports, are the same as in a
  single-antenna system. This is so because the distortion per antenna
  depends only on the input signal and has no dependence on intended
  spatial direction.
\item The EVM and ACLR values measured \emph{at the receiver} will
  also be the same as in the single-antenna case. This is so because
  the useful signal and the distortion receive the same array gain.
\item Neither the useful signal, nor the distortion, receives any
  array gain in spatial directions different from that defined by
  $\tau$. Over-the-air measurements in other directions will show
  substantially smaller received distortion power than in the
  direction of the intended receiver.
\end{itemize}
 
\section{Multi-user beamforming case}\label{sec:multiuser}

Next, consider again the two-antenna model and the case that the two
sinusoids that constitute the desired signal are destined to two
\emph{different} spatial directions, each direction corresponding to
one receiver. Let these spatial directions be defined in terms of the
two delays $\tau_1$ respectively $\tau_2$, both defined analogously to
$\tau$ above.  The conditions for constructive combination of the
desired signal are now
\begin{align}
 \phi_1^2-\phi_1^1 & = \omega_1\tau_1, \\
 \phi_2^2-\phi_2^1 & = \omega_2\tau_2.
\end{align}
In this case the distortion terms (\ref{eq:d1}) respectively (\ref{eq:d2}) add up
coherently in the directions defined by
\begin{align}
\tilde\tau_1 & = \frac{  (2\phi_2^2-\phi_1^2) - (2\phi_2^1-\phi_1^1)}{  2\omega_2-\omega_1} \\
\mbox{and}\qquad  \tilde\tau_2 & = \frac{  (\phi_2^2-2\phi_1^2) - (\phi_2^1-2\phi_1^1)}{  \omega_2-2\omega_1} .
\end{align}
In general, it holds that
\begin{align}
\tilde \tau_1 & \neq  \tau_1,  & \tilde \tau_1 & \neq  \tau_2 \\
\tilde \tau_2 & \neq  \tau_1,  & \tilde \tau_2 & \neq  \tau_2 .
\end{align}
Consequently, the distortion does add up constructively in distinct spatial directions,
but these directions are different from those of the two intended receivers.

Here the (transmitter) EVM and ACLR measured per antenna port are the same as in the
single-antenna case, and the same as in Section~\ref{sec:SU}.  In
contrast, the EVM and ACLR measured \emph{at the receiver} are
better. This is so because the useful signal receives the maximum
possible array gain (3 dB with two antennas), whereas the distortion
does not.  There will, however, be other directions into which the
distortion receives  array gain and over-the-air measurements
in those directions will show larger received distortion power.
 
\section{Discussion of Recent Literature}\label{sec:comm}

Much recent literature has been concerned with modeling the effects of
the distortion that arises from hardware impairments such as
non-linearities, especially in the Massive MIMO context.  For example,
\cite{bjornson2014massive} argues that ``the aggregate residual
transceiver impairments in the hardware'' can be modeled by additive
distortion noise terms that are statistically independent among the
antennas.  The same model is subsequently used in, for example,
\cite{zhang2015impact,papazafeiropoulos2016downlink,zhu2017analysis}.
If the distortion noise terms were statistically independent among the
antennas, the distortion would effectively be radiated
omni-directionally from the array. To understand why, consider a
fictitious receiver located in the electrical far-field of the
array. This receiver would see a sum of statistically independent
distortion noise terms that add up over the air. 
Only the sum of the power of these noises matters, and
this sum power only depends on the distance to the array -- but not on
the array geometry, nor on the relative array-to-receiver
geometry. But, as demonstrated in
Sections~\ref{sec:SU}--\ref{sec:multiuser}, the hardware
non-linearities lead to spatially directive emissions of
distortion. Hence, in conclusion, the distortion terms are correlated
between the antennas, which means that the model of
\cite{bjornson2014massive} is physically inaccurate.

Importantly, whether or not any ``appropriate compensation algorithms
have been applied'' such that the focus can be on ``residual''
hardware impairments (an assumption made in
\cite{bjornson2014massive}) appears to be irrelevant. If such compensation
results in the complete removal of the distortion, the entire issue is
immaterial. Conversely, if not, then the analysis in
Sections~\ref{sec:SU}--\ref{sec:multiuser} applies; effectively, a
model such as (\ref{eq:thirdorder}) then captures the characteristics
of the ``compensated'' hardware components.

As suggested in \cite{gustavsson2014impact}, the final result in terms
of EVM \emph{within the band} may in certain cases be the same when
comparing a physically correct behavioral model of the
non-linearities, and the independent-noise model of
\cite{bjornson2014massive}. Yet, in many cases -- and in particular,
if applied to the analysis of  single-user
beamforming, the assumption of independent distortion terms leads to
incorrect conclusions.

We next discuss out-of-band emissions more specifically.  This topic
has been the subject of many recent academic papers, but
unfortunately, the conclusions in several of these papers are unclear
or incorrect. This observation is important, given the current debate
on out-of-band emissions in 5G standardization and regulatory
discussions \cite{FCC}. Space limitations permit only short comments:
\begin{itemize}
\item The paper \cite{mollen2016out} declares that (under assumptions
  specified therein) ``... the absolute amount of disturbing power a
  victim that operates in an adjacent band suffers from is also
  reduced in the massive MIMO system, even if the ACLR in the
  single-antenna system and the MIMO-ACLR in the massive MIMO system
  are the same.'' But as shown above, in the single-user case, the
  out-of-band radiation is beamformed into the same direction as the
  useful signal, hence receiving the same array gain. This means that
  ``the absolute amount of disturbing power'' is not necessarily
  reduced.
  
\item More seriously, the paper \cite{blandino2017analysis} states
  that ``We show analytically that [out-of-band] OOB does not recombine and the
  array gain is experienced only inside the desired bandwidth.''  But
  as the above analysis demonstrates, constructive recombination of
  the distortion can occur regardless of $\omega_1$ and $\omega_2$.
  In particular, out-of-band radiation may recombine constructively
  even in the same direction as does the useful signal
  (Section~\ref{sec:SU}).  Hence, the cited claim from
  \cite{blandino2017analysis} is correct only under very specific
  conditions, which do not hold in general.

\item The paper \cite{zou2015impact} claims in its abstract that ``the
  impact of non-linear PAs and the resulting linear and non-linear
  multi-user interference, quantified in terms of the received
  signal-to-interference-plus-noise ratio (SINR), is largely dependent
  on the effective or observable linear gain in the user equipment
  (UE) receiver demodulation stage.''  But this statement is
  inaccurate.  The observable gain of the desired signal at the
  receiver is essentially the array gain; however, the amount of
  received out-of-band radiation is not a direct function of this
  gain. In fact, as illustrated in Section~\ref{sec:multiuser}, in a
  multi-user scenario, the out-of-band radiation generally is
  beamformed into other directions than is the useful signal.  In this
  case, the array gains of the useful signal and that of the
  distortion are different; so are the observable linear gains of
  these signals.
\end{itemize}

\section{Conclusion}\label{sec:concl}

In a multiple-antenna transmitter, the distortion that results from
hardware non-linearities is correlated between the antennas.  This
stands in contradiction to hardware impairment models popularized in
the communication theory literature \cite{bjornson2014massive}, and
which postulate that distortion is statistically independent among the
antennas. 

This observation has several implications:
\begin{itemize}
\item In the single-user beamforming case, the distortion resulting
  from non-linearities is beamformed into the same direction as the
  desired signal.

\item In the multi-user beamforming case, the distortion resulting
  from non-linearities is beamformed into distinct spatial directions,
  which are different from those of the desired signals.
\end{itemize}

Out-of-band interference may be the most serious form of distortion.
Given that this interference is beamformed into specific directions,
per-antenna ACLR constraints might be over-conservative in many
cases. An alternative may be to stipulate ACLR constraints on the
signal measured over-the-air at the intended receiver
location. However, such a constraint would not guarantee with
certainty that no other locations observe strong out-of-band
interference.
  
When finalizing this letter it was brought to our attention that some
of these conclusions have already been experimentally validated:
\cite{sienkiewicz2014spatially} demonstrated that out-of-band relative
power measurements taken at different points in space were different
from measurements taken at the antenna ports.

\section{Acknowledgment}
The authors would like to thank collaborators in the FP7-MAMMOET
project, and co-authors of
\cite{mollen2016out,blandino2017analysis,zou2015impact,bjornson2014massive,gustavsson2014impact},
for useful discussions on hardware impairments.

\end{document}